\documentclass[jgrga]{agutex}

\usepackage{graphicx}
\usepackage{amsmath}
\usepackage{amssymb}

\setkeys{Gin}{draft=false}

\authorrunninghead{OKA ET AL.}

\titlerunninghead{ELECTRON DIFFUSION REGION}

\authoraddr{Corresponding author: M. Oka,
Space Sciences Laboratory, University of California Berkeley, 7 Gauss Way, Berkeley, CA 94720-7450, USA.(moka@ssl.berkeley.edu)}

\begin{document}

%
%

\title{In-situ evidence of electron energization in the electron diffusion region of magnetotail reconnection}
%
%

%
%









\authors{M. Oka,\altaffilmark{1}, T.-D. Phan\altaffilmark{1}, M. {\O}ieroset\altaffilmark{1} and  V. Angelopoulos\altaffilmark{2}}

\altaffiltext{1}{Space Sciences Laboratory, University of California Berkeley, Berkeley, California, USA.}
\altaffiltext{2}{IGPP/ESS, University of California Los Angeles, California, USA.}

%
%


\begin{abstract}
Magnetic reconnection is an explosive energy-release process in laboratory, space and astrophysical plasmas. While magnetic fields can `break' and `reconnect' in a very small region called the electron diffusion region (EDR), there have been conflicting theories as to whether this region can be a place of rapid energization of plasmas. Here we report a fortuitous encounter of the EDR by THEMIS in the Earth's magnetotail where significant heating and demagnetization of electrons were observed. Additional energization was observed on both sides (immediate upstream and downstream) of the EDR, leading to a total of more than an order of magnitude energization across this region. The results demonstrate that, despite its minuscule size, the EDR does indeed contribute to the overall process of electron energization via magnetic reconnection.
\end{abstract}

%
%

%

\begin{article}

%
%

\section{Introduction}
Magnetic reconnection plays an important role during explosive energy-release phenomena in plasmas ranging from the Earth's magnetosphere to solar coronal and astrophysical applications. During reconnection, magnetic field lines of opposite directions `break' and `reconnect' in the diffusion region and the magnetic energy is quickly converted to particle energies. The diffusion region can have internal structures at ion-scale and electron-scale. The diffusion region at ion-scale is often called the Hall region and has been studied extensively \citep[e.g.][]{Sonnerup1979, Terasawa1983a, Hesse1999, Shay2001, Nagai2001a,Oieroset2001,Runov2003c,Borg2005,Eastwood2010a, Lu2010, Wang2014}. The diffusion region at electron-scale (or the `electron diffusion region' (EDR)) has also been studied extensively, as described below.

From theoretical point of view, it has been argued that electrons can be energized significantly in the EDR \citep[e.g.][]{Pritchett2006a, Fu2006}. Recent PIC simulations of the EDR revealed a fine structure called `striation' in the electron velocity distribution \citep{Ng2011,Ng2012, Bessho2015} which can further evolve into `swirls', `arcs' and `rings' \citep{Bessho2015,Shuster2015}. These fine structures are attributed to the particle meandering motion within the electron current layer and the energization (heating) mechanism can be attributed to the direct acceleration by the reconnection electric field. Furthermore, there can be additional energization immediately upstream of the EDR \citep[e.g.][]{Hoshino2005, Egedal2005, Egedal2010a, Chen2008}. It has been argued that the incoming flux of the magnetic fields expands as it approaches the reconnection point. Associated with this is a development of electron parallel anisotropy as well as electron energization by the parallel electric field \citep[e.g.][]{Egedal2005, Egedal2008, Egedal2013}.

On the other hand, theories predict that electrons can be energized by other reconnection processes such as slow shocks \citep{Petschek1984, Tsuneta1998a} and magnetic islands (or fluxropes in 3D) \citep[e.g.,][]{Drake2006, Fu2006, Wang2010a, Oka2010, Oka2010b, Tanaka2010}. Thus, it remains unclear how much the EDR contributes to the overall process of electron energization via magnetic reconnection.

A challenge, from observational point of view, is that the EDR is extremely small to be fully examined \citep[e.g.,][]{Shay2007, Hesse2014}. In the case of the Earth's magnetotail, a standard 2D picture of magnetic reconnection predicts that the length $\lambda_{\rm x}$ of the electron diffusion region (EDR) is 100-500 km in typical magnetotail plasma conditions at 25 R$_{\rm E}$ \citep{Hesse2014}. This is much smaller than the typical scale of the Earth's magnetotail, which is long (a few hundreds of $R_{\rm E}$ where $R_{\rm E}\sim$6371 km is the Earth’s radius) though narrow ($\sim$40 $R_{\rm E}$). Thus, there is probably not much chance for a spacecraft to encounter the EDR. Furthermore, considering the speed of the EDR motion due to a dynamical evolution of the magnetotail (typically $\sim$100 km/s), high-time resolution of plasma measurements are desired for a detailed study of the EDR.  For example, a 500 km long EDR moving at 100 km/s can be resolved with a measurement with the time resolution less than 5s.

Nevertheless, there have been cases of EDR detection based on (1) decoupling of ion and electron bulk flow velocities \citep{Nagai2011, Nagai2013} or (2) a higher-order scalar measure derived from particle data \citep{Scudder2012, Zenitani2012a}. The scalar measure can be the degree of electron non-gyrotropy (i.e., Agyrotropy $A\Phi_e$ of \cite{Scudder2012}) or the degree of energy dissipation (i.e., the dissipation measure $D_e^*$ of \cite{Zenitani2012a}). While one study reported energization weaker than that in the EDR downstream \citep{Nagai2013}, another study reported a strong temperature increase (by a factor of 2.5) \citep{Scudder2012}.

The purpose of this paper is to report yet another case of EDR detection in the Earth's magnetotail and to show evidence of substantial electron energization within the EDR (by a factor of $\sim$2 temperature increase). The observation was made by THEMIS in the Earth's magnetotail, and its high-quality particle data allowed us to examine electron distributions and energy spectra across the EDR.  With additional energization on both sides (immediate upstream and downstream) of the EDR, a total of more than one order of magnitude energization was observed across the EDR, demonstrating that, despite its minuscule size, the EDR does indeed contribute to the overall process of electron energization via magnetic reconnection.



\section{Observation}

We first present a brief description of the THEMIS instrumentation and the dataset (Section \ref{sec:instr}), followed by an overview and large scale contexts of the observation, in particular the extent of the Hall region (Section \ref{sec:overview}). We then describe the encounter of the electron diffusion region (EDR) on the basis of the observed time profiles of various plasma parameters (Section \ref{sec:encounter}). To establish the detection of the EDR, we examine pitch angle distributions (Section \ref{sec:pad}) as well as velocity distribution functions across the EDR encounter (Section \ref{sec:nongyro}). Finally, the electron energization is discussed quantitatively based on electron energy spectra (Section \ref{sec:heating}).

%
%
\begin{figure}[t]
\begin{center}
\includegraphics[width=20pc]{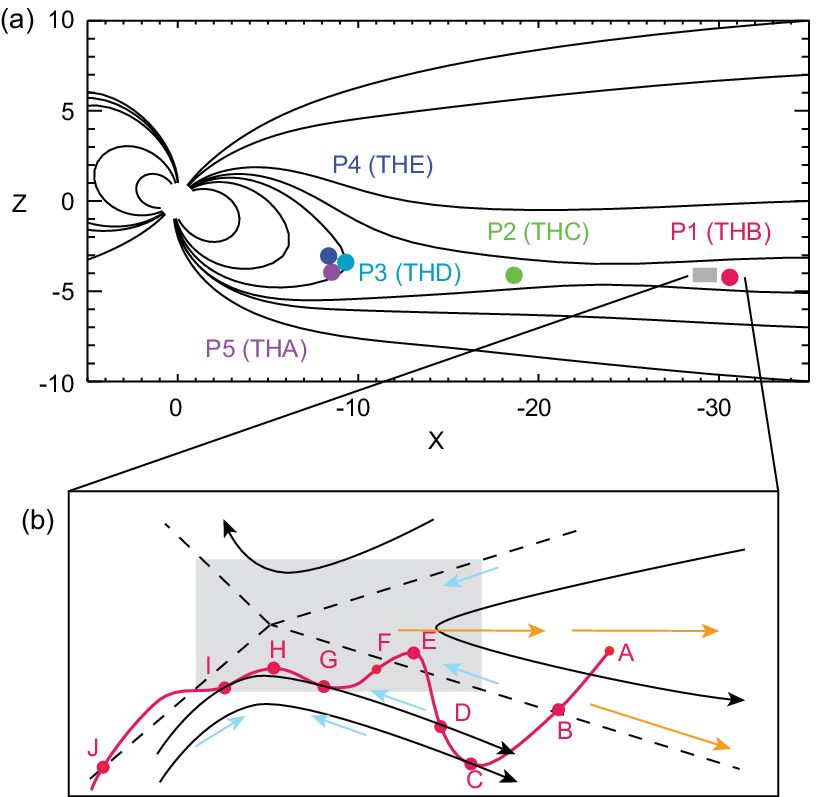}
\label{fig:orbit}
\end{center}
\end{figure}

\begin{figure}[t]
\begin{center}
\caption{Locations of the THEMIS spacecraft. (a) Locations of all five THEMIS spacecraft at 04:06 UT in the magnetotail. The black lines indicate the T-96 model of the magnetic field lines \citep{Tsyganenko1995a}.  (b) Schematic illustration of the inferred P1 trajectory (magenta) with respect to a 2D picture of the EDR. The light-blue and orange arrows indicate incoming and outgoing electrons, respectively.}
\label{fig:orbit}
\end{center}
\end{figure}

%
%
\begin{figure}[t]
\begin{center}
\includegraphics[width=20pc]{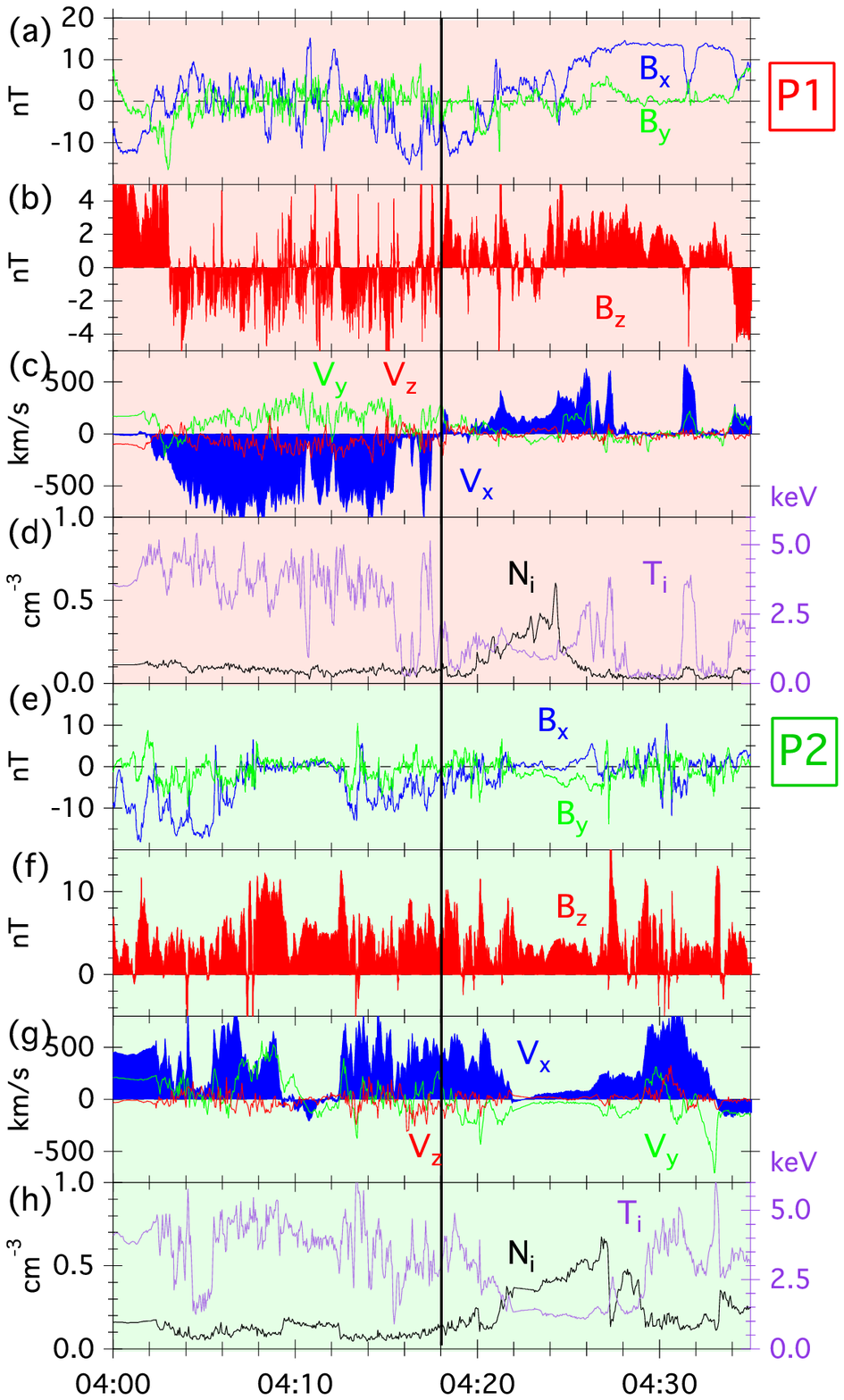}

\label{fig:overview}
\end{center}
\end{figure}

\begin{figure}[t]
\begin{center}

\caption{Overview and large-scale contexts of the THEMIS magnetotail reconnection event on 2009 February 7. (a-d) Time variations of plasma parameters observed by THEMIS P1. (e-f) The same parameters but for THEMIS P2.}
\label{fig:overview}
\end{center}
\end{figure}

%
%
\begin{figure}[t]
\begin{center}
\includegraphics[width=20pc]{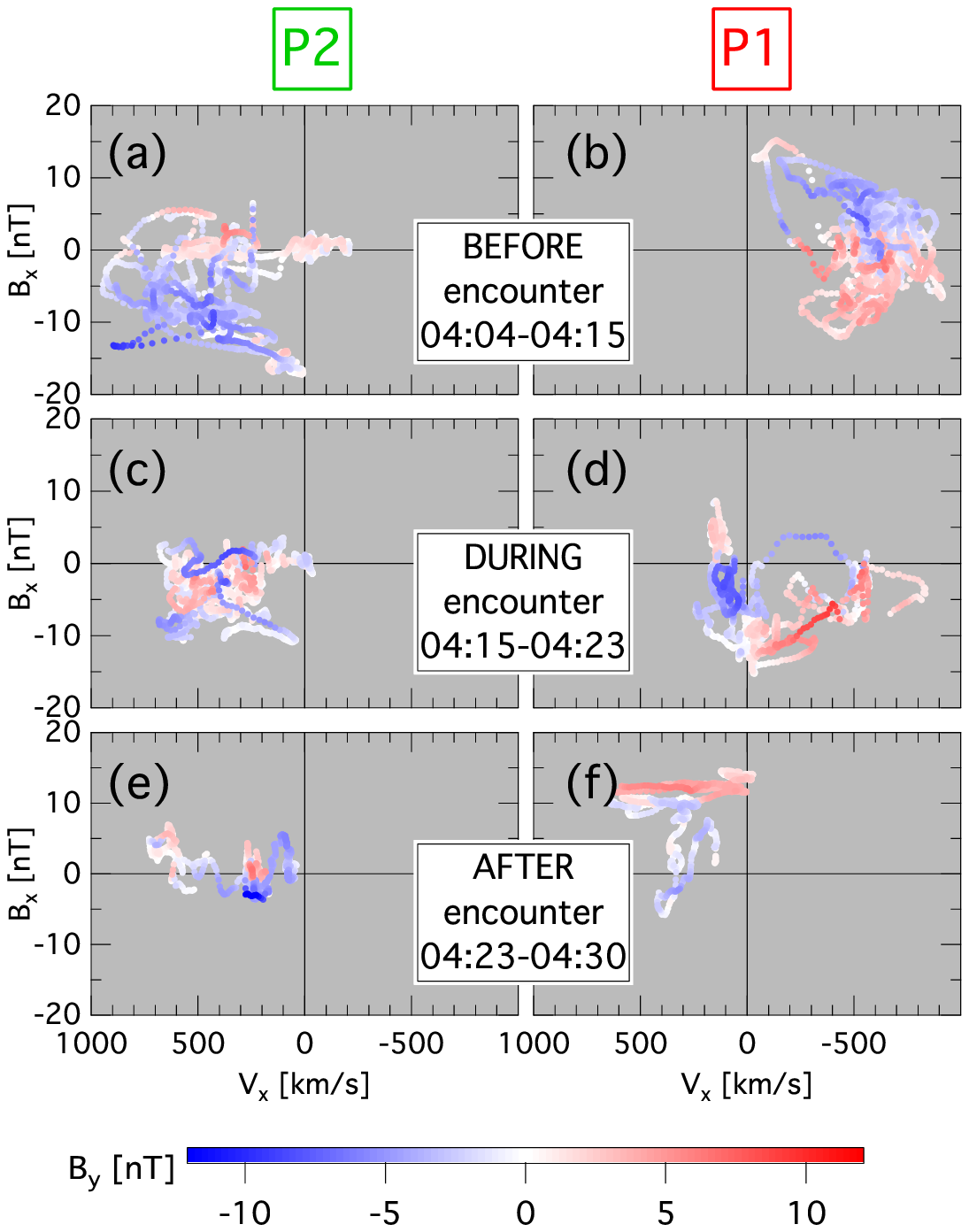}
\label{fig:hall}
\end{center}
\end{figure}

\begin{figure}[t]
\begin{center}
\caption{The quadrupole Hall structure of around the time of EDR encounter. $B_{\rm y}$ values are shown in the $V_{\rm x}$-$B_{\rm y}$ space obtained both before (uppper panels), during (center panels) and after (bottom panels) the encounter by P1 (right panels) and P2 (left panels). }
\label{fig:hall}
\end{center}
\end{figure}

\subsection{Instrumentation and Data}
\label{sec:instr}

We use data from the THEMIS mission \citep{Angelopoulos2009a}, which consists of five identical spacecraft P1-P5. For our observation of the magnetotail, we particularly focus on data with the highest resolution (the burst-mode data) from the P1 spacecraft. For ion and electron distributions, we use data obtained by the ESA instrument in the 0.03-28 keV energy range. For ion and electron moments, we combined ESA with SST, which measures ions and electrons in the higher energy ($>$32 keV) range. We confirmed that the ion and electron densities are consistent with each other during the duration of our observation. For magnetic and electric fields, we use data measured by FGM and EFI, respectively. Also, we use the Geocentric Solar Magnetospheric (GSM) coordinate system throughout the analysis unless otherwise noted.

\subsection{Large scale contexts}
\label{sec:overview}

The event was obtained on 2009 February 7. All five THEMIS probes (P1 - P5) were aligned in the sun-earth direction near the mid-night sector, allowing the study of the global evolution of the magnetotail associated with reconnection (Figure \ref{fig:orbit}(a)). It is already reported that magnetic reconnection started somewhere between P1 and P2 and that it moved tailward in association with a pressure increase at the locations of P3, P4 and P5 \citep{oka2011}. While P2 remained on the earthward side of the reconnection, P1 detected the tailward passage of the X-line. Thus, in this paper, we will focus on data from P1 in the vicinity of the X-line (diffusion region), although we also use data from P2 to retain informatoin on the large scale context of the investigation.

Figure \ref{fig:overview} shows an overview of the P1 and P2 data. This combined  overview provides large-scale contexts for the observation of the EDR  examined later in more detail. Figure \ref{fig:overview}(a-d) shows the time variations of plasma parameters for a 35 minute period during which the X-line passed by P1. From 04:02 UT, P1 stayed in the current sheet as evidenced by $B_{\rm x}$ fluctuating around 0 nT as well as the relatively high ion temperature $T_{\rm i}$ at around 3-5 keV. The plasma flow $V_{\rm i, x}$ was fast ($|V_{\rm i, x}|$ as large as $\sim$ 900 km/s) and directed tailward ($V_{\rm i, x} < 0$) whereas the magnetic field was directed southward ($B_{\rm z} < 0$), indicating that P1 was in the tailward side of the reconnection X-line. At 04:18 UT (as marked by the vertical solid line), the plasma flow turned earthward ($V_{\rm i,x} > 0$) and the magnetic field turned northward ($B_{\rm z}> 0$). The correlated reversals of both $V_{\rm i,x}$ and $B_{\rm z}$ indicate that the reconnection X-line passed by the P1 spacecraft. From 04:28 UT, $B_{\rm x}$ remained high at around 16 nT and the temperature remained low below 500 eV, indicating that P1 entered the lobe region, although there were tentative re-approaches to the current sheet at $\sim$04:31:45 UT and $\sim$04:34:15 UT.

Figure \ref{fig:overview}(e-h) shows the same plasma parameters but for P2. During most of the 35 minute period, $|B_{\rm x}|$ remained small, less than 10 nT, and $T_{\rm i}$ remained high, above 2.5 keV, indicating that P2 stayed in the current sheet throughout the observation. The magnetic field was northward ($B_{\rm z} > 0$) and the plasma flow was earthward ($V_{\rm i, x} < 0$), indicating P2 remained on the earthward side of the X-line. At $\sim$04:10 UT and $\sim$04:12 UT, there were flow reversals with no $B_{\rm z}$ reversal ($B_{\rm z}$ remained $>$2 nT), indicating P2 went outside the region of bursty bulk flows (BBF) \citep[e.g.][]{Angelopoulos1992} or the BBF itself diminished temporarily.

Figure \ref{fig:hall} shows the out-of-plane component of the reconnection magnetic field $B_{\rm y}$ in the $V_{\rm i, x}$-$B_{\rm x}$ space, demonstrating a spatial extent of the quadrupole Hall structure. For presentation purposes we excluded data points when a significant but transient bipolar structure (possibly a magnetic flux rope) was identified in the $B_{\rm z}$ time profile. Such a structure with an enhanced $B_{\rm y}$ core field was found during 04:16:54 - 04:17:00 UT and 04:21:00 - 04:22:00 UT in the P1 data and during 04:13:20 - 04:13:35 UT in the P2 data. We confirmed that the features presented below remain evident even when these data points were included and  our conclusion described below is unaffected.

During 04:04 - 04:15 UT, well before the X-line encounter by P1 at 04:18 UT, a quadrupole structure is evident when the scatter plots from both P1 (Figure \ref{fig:hall}(b)) and P2 (Figure \ref{fig:hall}(a)) are combined, indicating that the Hall region were extended more than the distance between P1 and P2 along $X_{\rm GSM}$, i.e., 12 R$_{\rm E}$ where $R_{\rm E}$ is the Earth's radius. This distance corresponds to $\sim$ 75 $d_{\rm i}$  where $d_{\rm i}$ is the ion inertia length. For the estimation, we used ion density measured in the lobe, $N_{\rm i}$ = 0.05 cm$^{-3}$, consistent with a typical value 0.05 cm$^{-3}$ in the lobe \citep{Pedersen2001,Pedersen2008}. Note again that the X-line was located somewhere between the two probes P1 and P2. If we assume the X-line was located in the middle of the two probes, then the size of the Hall region (when measured from the X-line) could have been $\sim$35 $d_{\rm i}$. Such a large spatial extent of the Hall region would be consistent with the extended electron jet observed in PIC simulations \citep[e.g.][]{Daughton2006b, Fujimoto2006c, Shay2007, Karimabadi2007} as well as the extended electron current layer seen in the magnetosheath \citep{Phan2007a} and in the magnetotail \citep[e.g.][]{Chen2008}.

During 04:15 - 04:23 UT, immediately before and after the X-line passage at 04:18 UT, the quadrupole Hall structure can be identified from P1 data alone (Figure \ref{fig:hall}(d)), indicating that the quadrupole Hall fields passed by P1. The scatter plot from P2 (Figure \ref{fig:hall}(c)), however, show a mixture of both positive and negative $B_{\rm y}$ in the quadrant of $V_{\rm x}>0$ and $B_{\rm x}<0$, indicating that the Hall structure was not extended to the P2 location at that time. We interpret this as evidence that the entire Hall region had moved tailward in association with the X-line retreat \citep{oka2011} and/or it had become shorter than $\sim$75$d_{\rm i}$.

During 04:23 - 04:30 UT, i.e. after the X-line encounter by P1 at 04:18 UT, P1 did not show the quadrupole feature (Figure \ref{fig:hall}(f)), suggesting that P1 exited the Hall region soon after the X-line encounter. Note that P1 quickly entered the lobe region from $\sim$04:25 UT so that most of the data points in Figure \ref{fig:hall}(f) are clustered above $B_{\rm x} >$ 10 nT.

We also note that WIND measurements around the time of this event (with the time shift) show that the interplanetary magnetic field (IMF) was $B_{\rm x} \sim$ 1.5 nT, $B_{\rm y} \sim$ -2 nT and $B_{\rm z} \sim$ -2 nT, solar wind speed was $\sim$320 km/s and the plasma pressure was $\sim$1.3 nPa. Such a solar wind condition may be providing a condition of slow convection, expanded magnetotail cross section and a slow reconnection process. Because a finite $B_{\rm y}$ value can twist the magnetotail \citep[e.g.][]{Sibeck1985, Tsyganenko2004}, the $B_{\rm y}$ component in Figure \ref{fig:hall} could be due to a possible twist of the magnetotail. However, we confirmed that the quadrupole feature is not sensitive to a $B_{\rm y}$ offset of 2 nT.

%
%

\begin{figure}[t]
\begin{center}
\includegraphics[width=20pc]{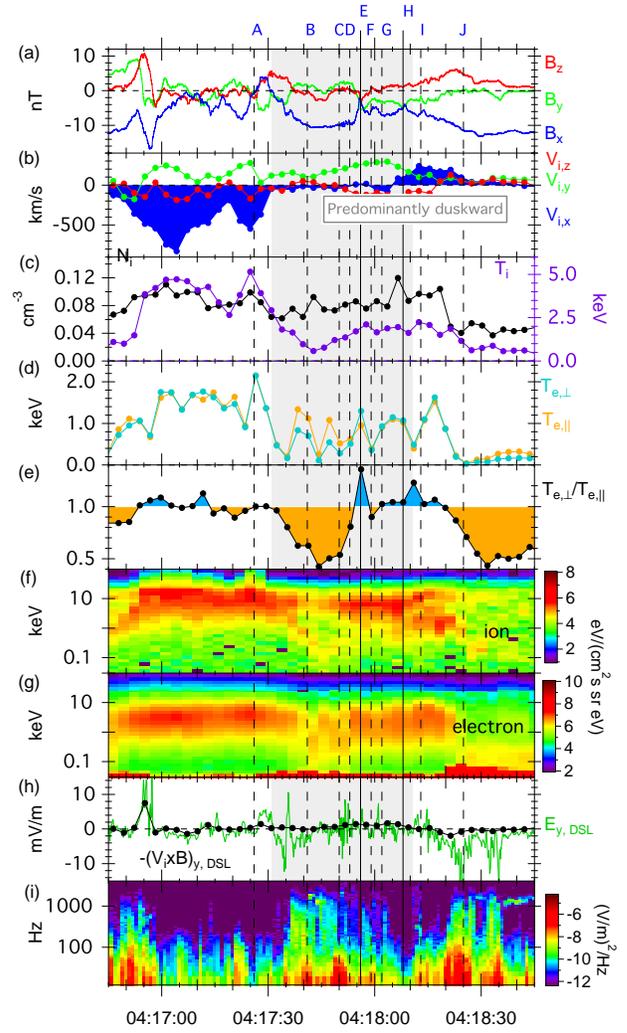}
\caption{The observed time profiles around the time of X-line passage by P1, which were used to reconstruct the inferred trajectory in Figure \ref{fig:orbit}(b).}
\label{fig:blowup}
\end{center}
\end{figure}

\subsection{Encounter with the electron diffusion region}
\label{sec:encounter}

We now examine in more detail the data at and around the time of X-line passage by P1, i.e. 04:18 UT, and look for an evidence of electron diffusion region (EDR). Our observations (as described below) suggest that P1 wandered across the EDR and we infer the schematic picture of Figure \ref{fig:orbit}(b) for the P1 trajectory with respect to a 2D picture of the EDR.


Figure \ref{fig:blowup} shows an expanded view of time profiles of various plasma parameters from a 2-minute period around the time of the X-line passage, 04:18 UT, serving as a basis of the reconstruction illustrated in Figure \ref{fig:orbit}(b).

First, P1 was on the tailward side of the X-line, as indicated by the tailward ($V_{\rm i,x} < 0$) jet. Also, P1 remained close to the current sheet center until about 04:17:35 UT, as indicated by $B_{\rm x}$ fluctuating around 0 nT, a relatively high density ($N_{\rm i}\sim$0.08 cm$^{-3}$) and high temperatures ($T_{\rm i}\sim$3 keV, $T_{\rm e} \sim$ 1.6 keV)). At time A, 04:17:26 UT (as indicated by Point A in the schematic illustration as well as the first vertical dashed line from left in the time profiles), P1  was in the downstream region but right at the current sheet center as evidenced by $B_{\rm x} \sim$ 0 nT.

Then, during the next $\sim$40s, 04:17:35 - 04:18:15 UT (highlighted by the gray background), P1 stayed within the key region of the reconnection as indicated by the ion flow directed predominantly duskward ($|V_{\rm i,y}| > |V_{\rm i,x}|, |V_{\rm i,z}|$).  $|B_{\rm x}|$ stayed relatively high at $\sim$ 10 nT but ion density $N_{\rm i}$ also remained high at  $\sim$0.08 cm$^{-3}$, suggesting that P1 was not too far away from the current sheet center. In fact, both cold and hot ion components co-existed there especially in the first half of the key region 04:17:35 - 04:17:53 UT (Figure \ref{fig:blowup}(f,g)), indicating that the spacecraft was at the boundary between the cold upstream region and the hot plasma sheet. The electron parallel temperature $T_{\rm e||}$ was much larger than the electron perpendicular temperature $T_{\rm e\perp}$ with $T_{\rm e||}/T_{\rm e\perp} > 1.2$ (Figure \ref{fig:blowup}(d,e)).  Also, the parallel anisotropy was associated with an enhancement of broadband electrostatic noise (Figure \ref{fig:blowup}(i)). These features are generally consistent with reconnection separatrices \citep[e.g.][]{Nagai2001a, Drake2003, Fujimoto2014}. A  study with PIC simulations reported that interactions between incoming cold electrons and outgoing hot electrons lead to enhanced wave activities along reconnection separatrix \citep{Fujimoto2014}. 

Nevertheless, P1 temporarily approached the current sheet center at time E ($\sim$04:17:56 UT) and time H ($\sim$04:18:08 UT) as indicated by the abrupt decrease of $|B_{\rm x}|$ down to 2.0  nT and 3.8 nT, respectively. The perpendicular temperature $T_{\rm e, \perp}$ was comparable to or higher than the parallel temperature $T_{\rm e, ||}$ at and around these two times (Figure \ref{fig:blowup}(e), times E - I). Such a perpendicular heating has been predicted by simulations of the EDR \citep[e.g.,][]{Pritchett2006a, Bessho2015, Ng2011, Ng2012}. Thus, the observed perpendicular heating can be regarded as an indication of EDR detection.


The features of perpendicular heating described above already imply that P1 encountered the EDR. To establish the detection of the EDR, we further examine pitch angle distributions as well as full-3D velocity distributions of electrons in the following subsections.

\subsection{Pitch angle distributions}
\label{sec:pad}

\begin{figure*}[t]
\begin{center}
\includegraphics[width=41pc]{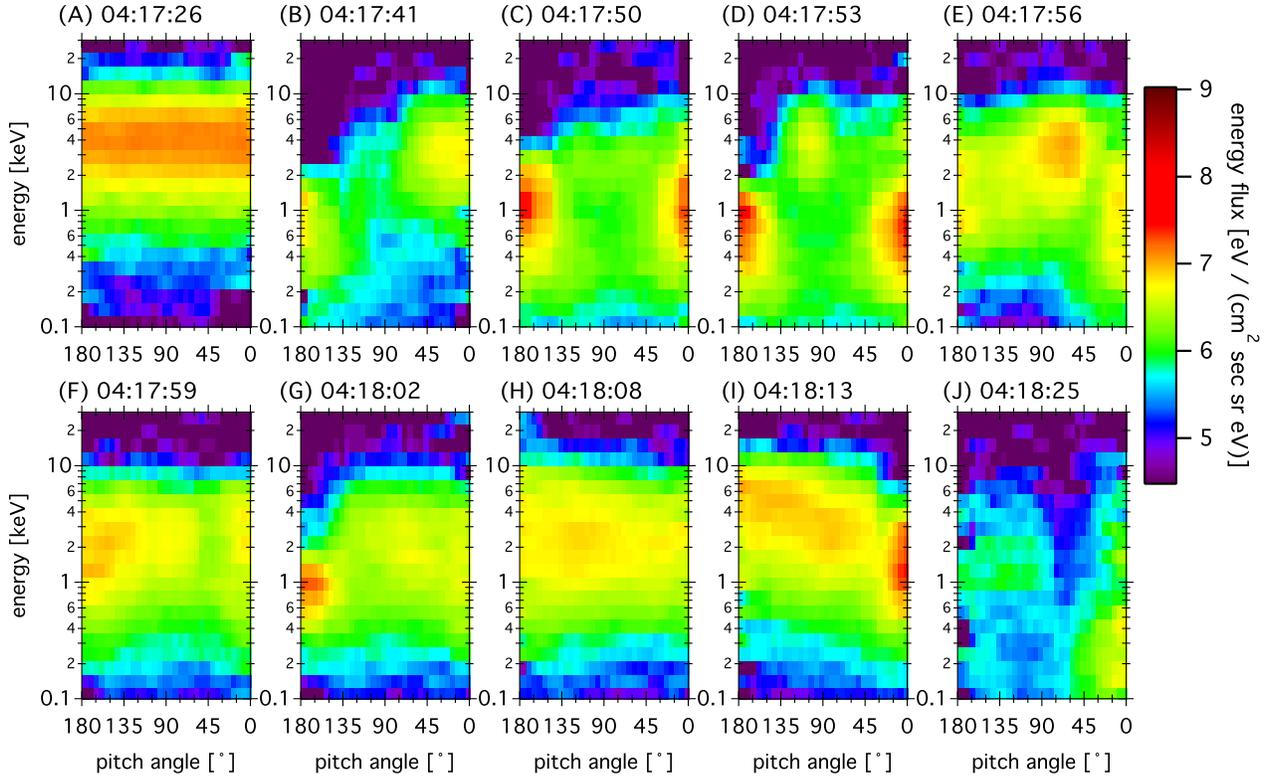}
\caption{Time evolution of electron pitch angle distributions. The times are chosen to match times A - J of Figure \ref{fig:blowup}. The horizontal axis ranges from 180$^{\rm o}$ pitch angle to 0$^{\rm o}$ pitch angle because the P1 spacecraft was in the southern hemisphere and the magnetic field line was directed from left to right in the schematic illustration of Figure \ref{fig:blowup}(a).}
\label{fig:PAD}
\end{center}
\end{figure*}

Figure \ref{fig:PAD} shows the evolution of pitch angle distributions of electron energy flux at times A - J, demonstrating the key identification of the perpendicular heating. 

At time A, electrons were energetic with a largest energy flux at $\sim$4 keV and showed an isotropic pitch angle distribution, indicating that P1 was in the reconnection downstream. At time B, bi-directional, field-aligned electron beams were observed. The incoming beam at $\sim$180$^{\rm o}$ was carrying lower energies ($\sim$ 1 keV) whereas the outgoing beam at $\sim$0$^{\rm o}$ was carrying higher energies ($\sim$ 3 keV). Such a bi-directional distribution with asymmetry in energy indicates that P1 was at the reconnection separatrix with an intensified Hall current \citep[e.g.][]{Fujimoto1997a, Nagai2001a, Manapat2006} and that the magnetic field line was connected to a heating site (or the EDR). At time C, P1 observed a bi-directional distribution again but the peak energy was symmetric at $\sim$ 1 keV indicating that the magnetic field line was not connected to the heating site (or the EDR). Nevertheless, the observed energy $\sim$ 1 keV is still significantly larger than the electron energy of $\lesssim$ 100 eV in the upstream region observed after 04:18:25 UT (Figure \ref{fig:blowup}(h)), indicating that there already exists an energization process at this location. Note that a typical electron temperature in the lobe region is $\lesssim$100 eV range \citep[e.g.][]{Pedersen2008}. The energization up to $\sim$ 1 keV has been interpreted as a consequence of electron trapping in an expanding flux tube and associated parallel electric fields \citep[e.g.][]{Egedal2005}.
%
%

At time D,  a third component appeared at pitch angle $\sim$110$^{\rm o}$ and energy $\sim$ 5 keV in addition to the counter-streaming electrons. This third component became even more pronounced during the next sampling time time E, although the energy flux of the counter-streaming electrons had become less, which we interpret as the spacecraft entering deeper into the heating site. This time, 04:17:56 UT, was in fact when P1 temporarily approached the current sheet center as indicated by the sharp drop of $|B_x|$ to 2.0 nT.

While the distribution remained complex at the next sampling time F, the mildly ($\sim$ 1 keV) energetic electrons appeared again at time G, only in the $\sim$180$^{\rm o}$ direction. This indicates that the magnetic field was connecting the heating site to the immediate upstream region. Then, P1 observed another episode of perpendicular heating at time H (04:18:08 UT) which is the time of the second approach to the current sheet center with a $|B_{\rm x}|$ decrease down to 3.8 nT. Furthermore, at time I (04:18:13 UT), P1 again observed a distribution similar to the distribution at time G, although the direction of mildly-energetic electron flow was opposite and was $\sim$0$^{\rm o}$. The apparent reversal of the direction of electron streaming indicates that the X-line passed by P1 between 04:18:02 UT (Time G) and 04:18:13 UT (time I). Later on at time J (04:18:25 UT), P1 observed a Hall current feature similar to those at time B but with a reversed direction.

The perpendicular heating as observed at times D, E and H during the correlated reversals of $V_{\rm i,x}$ and $B_{\rm z}$ is consistent with P1 being in the EDR \citep[e.g.][]{Pritchett2006a, Ng2011,Ng2012, Bessho2015}.

%
%
\begin{figure*}[t]
\begin{center}
\includegraphics[width=36pc]{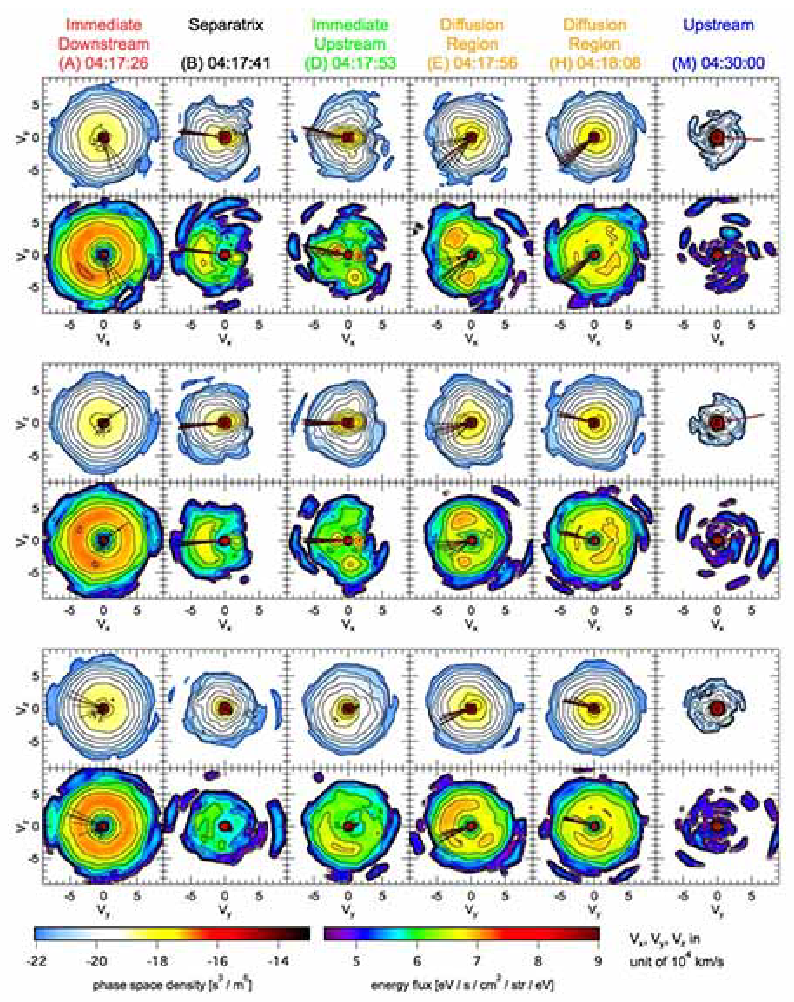}
\label{fig:nongyro2}
\end{center}
\end{figure*}
\begin{figure*}[t]
\begin{center}
\caption{A variety of electron velocity distributions during magnetotail reconnection on 2009 February 7. The images represent 2D slices of the 3D interpolated distributions in the $v_{\rm x}-v_{\rm y}$ plane (1st and 2nd rows), $v_{\rm x}-v_{\rm z}$ plane (3rd and 4th rows) and $v_{\rm y}-v_{\rm z}$ plane (5th and 6th rows). The velocity distributions of both phase space density (blue-yellow-red color scheme) and energy flux (rainbow color scheme) are shown. The black lines superposed in each image indicate magnetic field directions obtained every 0.25s. The red lines are their averages.}
\label{fig:nongyro2}
\end{center}
\end{figure*}

\subsection{Velocity distributions}
\label{sec:nongyro}

Figure \ref{fig:nongyro2} shows electron velocity distributions at six selected times, demonstrating the key identification of the electron non-gyrotropy in the EDR along with other distribution types. The first, second, third, fourth and fifth column from the left of this figure show distributions at times A, B, D, E and H, respectively, and are different representations of the data in panels A, B, D, E and H of Figure \ref{fig:PAD}. While phase space density (s$^3$/m$^6$) is a classical way of displaying electron distributions (as shown in the 1st, 3rd and 5th rows), we also use energy flux (eV/s/cm$^2$/str/eV) to visually emphasize the angular distribution of electrons (as shown in the 2nd, 4th and 6th rows). 

At time D, there is a relatively high-energy component in the $V_{\rm y}<0$ range in addition to the (relatively low energy) bi-directional components (See the top two panels in the third column of Figure \ref{fig:nongyro2}). This component corresponds to the previously mentioned third component at pitch angle $\sim$110$^{\rm o}$ and energy $\sim$5 keV in Figure \ref{fig:PAD}(D) and it exhibits a non-gyrotropic feature as described below.

In general, a non-gyrotropic distribution indicates demagnetization of electrons. If magnetized, electrons can complete their gyro-motion around a magnetic field line and would exhibit a symmetric velocity distribution when viewed in the plane perpendicular to the magnetic field.  The bottom two panels of the third column of Figure \ref{fig:nongyro2} (i.e., a slice along the $V_{\rm y}-V_{\rm z}$ plane) show such a slice because the magnetic field was parallel to the $V_{\rm x}$ axis of the velocity space. These bottom panels clearly indicate particles were moving perpendicular to the magnetic field (but not gyrating).

This non-gyrotropic component became even more pronounced during the next sampling time, E, at 04:17:56 UT, as the energy flux of the counter-streaming electrons decreased, which (as discussed earlier) we interpret as the spacecraft entering deeper into the heating site. As also argued earlier, this interpretation is corroborated by the fact that P1 experienced its closest approach to the current sheet center at that time, as indicated by the sharp drop of $|B_{\rm x}|$ to a minimum of 2.0 nT (Figure \ref{fig:blowup}(a)). The elongated component, as we have seen in time D, can still be identified but it is somewhat rotated in the $V_{\rm y}-V_{\rm z}$ plane. In addition to this primary component (in $V_{\rm y}<0$), a secondary component appeared as can be seen in the bottom panel of the third column (in $V_{\rm y}>0$). This secondary component makes the distribution a little closer to a gyrotropic distribution but not as much as what we can see in the fully gyrotropic distribution of the immediate downstream, displayed in the left most plane of the figure. Note also, the magnetic field direction has also changed since time D, but the three dimensional configuration of the velocity distribution is clearly non-gyrotropic. (We also examined a slice perpendicular to the magnetic field line by transforming from the GSE coordinate to the field-aligned coordinate, and the distribution looked very similar to the ones in the 3rd and 4th panel of the third column.)
A scenario of varying magnetic field direction with gyrotropic distribution cannot explain such a complex distribution. The origin of the complexity is unclear but it could be related to the complexity observed in recent high-resolution particle simulations \citep{Bessho2015}. 


The apparent non-gyrotropy at time E is an indication of electron demagnetization and reinforces our conclusion of the previous subsection that the spacecraft was in the EDR at time E. As for time D, the existence of the lower-energy, bi-directional components (usually present at the inflow region) indicates that the distribution was measured mostly in the immediate upstream region, although the apparent non-gyrotropy is likely to be due to the approach to the EDR. We speculate that this co-existence is due to the limited time resolution of electron measurement (3s) and/or different gyro-radius of electrons of different energies. Note that a higher-resolution (0.25s) measurement of magnetic field shows that the field was stable during this 3s sampling time, as shown by the black lines in Figure \ref{fig:nongyro2} (the red line shows the average direction).

P1 observed further increase in perpendicular heating at time H (04:18:08 UT), 4 spins later, at the time of a second approach to the current sheet center evidenced by a $|B_{\rm x}|$ decrease down to 3.8 nT. The distribution at time H, however, is not as clearly non-gyrotropic as at E and not as isotropic as at A (downstream). Nevertheless, before and after time H, there was a component of mildly ($\sim$ 1 keV) energetic, field-aligned electrons flowing tailward and earthward, respectively (Figure \ref{fig:PAD}(G, I)). The reversal at time H indicates that the EDR had passed by P1 at around time H. Thus, it appears P1 was within the EDR at time H despite the lower degree of non-gyrotropy.



\subsection{Energy spectra}
\label{sec:heating}

\begin{figure}[t]
\begin{center}
\includegraphics[width=20pc]{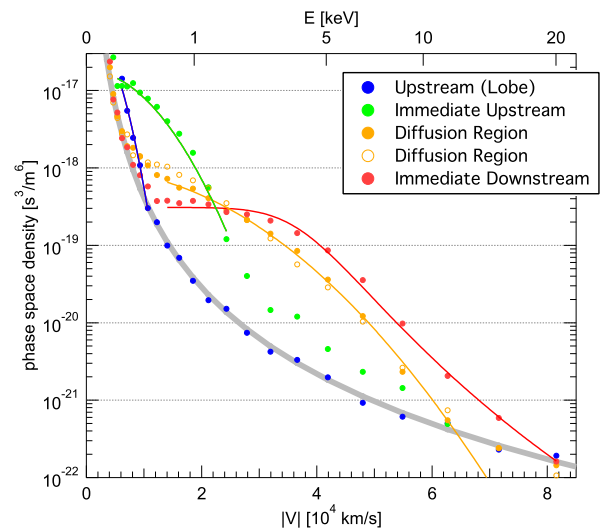}
\caption{Evolution of electron energy spectra across the EDR. The curves show 1D slices of the 3D distributions without interpolation but averaged over finite pitch angle ranges. Because the distributions were not isotropic (see text and Figures \ref{fig:PAD} and \ref{fig:nongyro2} for more details), we are showing 1D cuts in the direction of most prominent phase-space-density. For the immediate upstream region (green curve), the distribution $f(v_{||})$ is a parallel cut (i.e., an average over pitch angle ranges [0$^{\rm o}$, 35$^{\rm o}$] and [145$^{\rm o}$, 180$^{\rm o}$]. For all other curves, the distributions $f(v_{\perp})$ are perpendicular cuts (i.e., averages over a pitch angle range [65$^{\rm o}$, 115$^{\rm o}$]). The gray curve indicates the detection limit (i.e., one-count-level) of the ESA instrument.}
\label{fig:cut1d}
\end{center}
\end{figure}

Let us now examine the degree of heating (energization). Figure \ref{fig:cut1d} shows 1D slices of electron distributions from five selected regions, demonstrating the significance of electron energization within the EDR. In order to show the most prominent phase-space-density, we used perpendicular cuts for all cases except the case of immediate upstream (time D, colored green) in which a parallel cut was used (As already described in earlier sections, electrons were energized in the parallel direction in the immediate upstream).

In the upstream (lobe) region, the temperature was low and significant particle counts were recorded only in the lower velocity range $<10^4$ km/s. In general, it is difficult to evaluate temperatures in the lobe region, although a typical range of lobe temperatures is shown to be $\lesssim$100 eV \citep{Pedersen2001, Pedersen2008}.  In our case, the lobe temperature from the second order moment of a 1 min interval around 04:30 UT was $\sim$ 60 eV. A fit to the blue curve in Figure \ref{fig:cut1d} (taken at time 04:30:30 UT) resulted in the temperature of $\sim$ 60 eV. Please note again that this lobe region was selected from a time interval ($\sim$04:30 UT) far away from what is shown in Figure \ref{fig:overview}.

In the immediate upstream region, we found that the parallel temperatures from the second order moment at times C and D were $\sim$520 and $\sim$630 eV, respectively. The parallel temperature from a fit to the 1D cut shown in Figure \ref{fig:cut1d} (green curve taken from time D) was $\sim$350 eV. This value is somewhat smaller than those obtained from the moments but confirms the pre-heating before the EDR encounter.

In the EDR, the perpendicular temperatures from the second order moment at times E and H were $\sim$1.3 and $\sim$1.1 keV, respectively, indicating a further temperature increase within the EDR. A Maxwellian distribution can nicely fit the 1D perpendicular cut from time E (orange filled circles) and the derived temperature was $\sim$1.5 keV. 

The distribution became `flattop'-like with a non-thermal tail in the immediate downstream region \citep[e.g.][]{Asano2008, Chen2009, Egedal2010a, Wang2010b, Teh2012a, Nagai2013}. Here, we used time A (04:17:26 UT) to represent distributions of immediate downstream because $B_{\rm x}$ was small, $\sim$0.31 nT, indicating that the spacecraft was very close to the current sheet center. In general, a flat-top distribution with a non-thermal tail can be represented by the empirically derived, modified (flattened) Lorentzian \citep[e.g.][]{Feldman1982, Feldman1983,Thomsen1983a, Chateau1989}
\begin{equation}
f_{L}(v) = \frac{N_L \kappa \sin{(\pi/2\kappa)}}{\pi^2 v_{\perp L}^2 v_{||L}} \left[1+\left(\frac{v_{\perp}}{v_{\perp L}}\right)^{2\kappa}+\left(\frac{v_{||}}{v_{||L}}\right)^{2\kappa} \right]^{-\frac{\kappa+1}{\kappa}}
\end{equation}
where $N_L$ is the density, $\kappa$ is the spectral index and $v_L$ is the location and sharpness of the spectral break (or `shoulder'). In the higher energy limit ($v\gg v_L$), the distribution approaches a power-law $f \propto v^{-2(\kappa+1)}$ as is the case with the kappa distribution \citep[e.g.][]{Olbert1968, Vasyliunas1968a}. In the lower energy limit ($v\ll v_L$), the distribution becomes flat at $f = N_L \kappa \sin{(\pi/2\kappa)}/(\pi^2 v_{\perp L}^2 v_{||L})$. We found that the modified Lorentzian can best fit the perpendicular cut of the distribution in Figure \ref{fig:cut1d} with $N_L$ = 0.12$\pm$0.04 cm$^{-3}$, $v_{\perp L}$ = 3.9$\pm$0.5$\times$10$^{4}$ km/s and $\kappa$ = 4.1$\pm$1.0. The derived speed $v_L$ corresponds to the break energy 4.2 keV. The Maxwellian and kappa distributions could not fit the data as well. 

We note that there has been a physical explanation to the almost isotropic flat-top feature \citep[e.g.][]{Dum1978, Fujimoto2014, Egedal2015}. The entire spectral shape including both the flat-top and non-thermal tail features are also discussed from a statistical mechanics point of view with non-Euclidean metrics induced by $L_p$ norms \citep[See Eq.(66) of][]{Livadiotis2016}. 

Here, we evaluated the temperature of the flat-top distribution from the second order moment of the entire 3D distribution and it was $\sim$2.1 keV.  This value is higher by a factor of $\sim$1.8 than the temperature within the EDR and indicates that there was an additional energization in the immediate downstream region. Thus, through three different stages (i.e., immediate upstream with $T_e\sim$0.58 keV, EDR with $T_e\sim$1.2 keV and immediate downstream with $T_e\sim$2.1 keV), the electron temperature increased by a factor of $\sim$35 (from $\sim$60 eV to ~2.1 keV).  Such a significant energization is qualitatively consistent with PIC simulations of forced reconnection \citep[e.g.][]{Pritchett2006a}.

\section{Discussion and Conclusion}
\label{sec:discussion}

It appears that, on February 7, 2009, the THEMIS P1 spacecraft encountered the EDR because, during correlated reversals of $V_{\rm ix}$ and $B_{\rm z}$ in particular at times of approach to the current sheet center (i.e., at time of temporal decrease of $|B_{\rm x}|$), electrons were heated in the direction perpendicular to the local magnetic field. The non-gyrotropy of electrons were evident at least at time E (04:17:56 UT), indicating that electrons were, in fact, non-magnetized within the EDR.  However, the non-gyrotropy was less enhanced at time H, indicating that an EDR does not always show an enhanced non-gyrotropy. 

The perpendicular heating within the EDR was significant and the temperature reached $\sim$1.2 keV. Additional energization was observed on both sides (immediate upstream and downstream) of the EDR, leading to more than an order of magnitude energization across this region (from 60 eV in the lobe to $\sim$2.1 keV in the immediate downstream of the EDR). The results demonstrate that, despite its minuscule size, the EDR does indeed contribute to the overall process of electron energization via magnetic reconnection. 

However, the contribution of the EDR to the overall energization process, i.e., $\Delta T_{\rm EDR}/\Delta T_{\rm total}$ is about 30\%, where $\Delta T_{\rm EDR}$ ($\sim$0.6 keV) is the temperature difference between the EDR and immediate upstream, and $\Delta T_{\rm total}$ ($\sim$2 keV) is the total temperature increase across the EDR. On the other hand, the immediate downstream showed a temperature increase of about 0.9 keV and the contribution is about 40\%. Thus, the degree of energization within the EDR was slightly smaller than the additional energization in the immediate downstream. This conclusion is consistent with \cite{Nagai2013} who asserted that heating and acceleration are weak in the central intense current layer. Nevertheless, our observation demonstrates that electron energization across the EDR (including the immediate upstream and downstream regions) is significant.

Regarding the physics of EDR, the rate of reconnection $E_{\rm R}$ is an important parameter. In our case, the reconnection rate $E_{\rm R}$ is difficult to estimate because of large fluctuations in the electric field data. Throughout the observation, the y-component of the electric field data was available as shown in Figure \ref{fig:blowup}(h) in the Despun, Sun-pointing L-momentum vector (DSL) coordinate system. Here, the y-component in the DSL coordinate is the best measured component with minimum solar illumination effects.  During the time period of perpendicularly heated electrons (i.e., 04:17:50 - 04:18:05 UT), the standard deviation $\sigma$  = 2.39 mV/m was larger than the average $\overline{E_{\rm y, DSL}}$ = 0.419 mV/m, demonstrating the difficulty of measuring $E_{\rm R}$. We note, for reference, that the expected value of $E_{\rm R} (= \alpha V_A B_0) $ is $\sim$2 mV/m for an assumed reconnection rate $\alpha$ = 0.1 with an outflow speed (or $\sim$Alfv\'{e}n speed) $V_A$ = 1500 km/s and a lobe magnetic field $B_{\rm 0}$ = 15 nT and that E$_{\rm y, DSL}$should remain positive (i.e., duskward) for a steady state, magnetotail reconnection. The origin of the electric field fluctuation is unclear. It may be a part of the reconnection-induced turbulence \citep[e.g.][]{Eastwood2009} or waves \citep[e.g.][]{Fujimoto2014, Egedal2015}.

The length of the EDR in the outflow direction is also an important parameter because it can depend on the reconnection rate $E_{\rm R}$ \citep[e.g.][and references therein]{Hesse2014}. In our case study, it is again  difficult to estimate, partly because the observation was made by a single spacecraft and it is difficult to estimate the speed of EDR motion. If we assume EDR was retreating at a constant speed of $\sim$100 km/s \citep[e.g.][]{Baker2002} or $\sim$0.1 V$_{\rm A}$ \citep{Oka2008} (where $V_{\rm A}$ is the Alfv\'{e}n speed), the duration of EDR observation from time E to time H ($\sim$12s) corresponds to $\sim$1200 km. This is somewhat larger than the theoretically predicted values of 100 - 500 km \citep{Hesse2014}, suggesting that the prediction and/or observation need to be refined. Also, the size of the EDR may be different if the theory was expanded to three dimensions and/or if the observations were made by multiple spacecraft with higher time resolution. 

The Magnetospheric Multi-Scale (MMS) mission, with its ultra high-time resolution measurements, will be able to address the EDR size and how the heating and demagnetization are involved in the rate of reconnection. While our THEMIS event demonstrated that electrons can be non-gyrotropic and heated significantly at the EDR, we expect MMS to verify and establish the generality of electron heating and non-gyrotropic features at the EDR.

\begin{acknowledgments}
We thank Lynn B. Wilson III and George Livadiotis for drawing our attention to different representions of the flat-top distribution. This research was funded by NASA grant NNX08AO83G at UC Berkeley. We also acknowledge NASA contract NAS5-02099 and all members of the THEMIS mission for use of data. Specifically: J. W. Bonnell and F. S. Mozer for use of EFI data,  D. Larson and R. P. Lin for use of SST data, C. W. Carlson and J. P. McFadden for use of ESA data and A. Roux and O. LeContel for use of SCM data.  We also thank K. H. Glassmeier, U. Auster and W. Baumjohann for the use of FGM data provided under the lead of the Technical University of Braunschweig and with financial support through the German Ministry for Economy and Technology and the German Center for Aviation and Space (DLR) under contract 50 OC 0302. The THEMIS data as well as the software (SPEDAS) for retrieval and analysis of the data are freely available at the THEMIS website (http://themis.ssl.berkeley.edu).  
\end{acknowledgments}

%
%
%
%
%
\bibliographystyle{agufull08}
%
%
%
%


\begin{thebibliography}{68}
\providecommand{\natexlab}[1]{#1}
\expandafter\ifx\csname urlstyle\endcsname\relax
  \providecommand{\doi}[1]{doi:\discretionary{}{}{}#1}\else
  \providecommand{\doi}{doi:\discretionary{}{}{}\begingroup
  \urlstyle{rm}\Url}\fi

\bibitem[{\textit{Angelopoulos et~al.}(1992)\textit{Angelopoulos, Baumjohann,
  Kennel, Coroniti, Kivelson, Walker, and Paschmann}}]{Angelopoulos1992}
Angelopoulos, V., W.~Baumjohann, C.~F. Kennel, F.~V. Coroniti, M.~G. Kivelson,
  R.~J. Walker, and G.~Paschmann (1992), {Bursty Bulk Flows in the Inner
  Central Plasma Sheet flows}, \textit{J. Geophys. Res.}, \textit{97}(A4),
  4027--4039.

\bibitem[{\textit{Angelopoulos et~al.}(2008)\textit{Angelopoulos, McFadden,
  Larson, Carlson, Mende, Frey, Phan, Sibeck, Glassmeier, Auster, Donovan,
  Mann, Rae, Russell, Runov, Zhou, and Kepko}}]{Angelopoulos2009a}
Angelopoulos, V., J.~P. McFadden, D.~Larson, C.~W. Carlson, S.~B. Mende,
  H.~Frey, T.~D. Phan, D.~G. Sibeck, K.-H. Glassmeier, U.~Auster, E.~Donovan,
  I.~R. Mann, I.~J. Rae, C.~T. Russell, A.~Runov, X.-Z. Zhou, and L.~Kepko
  (2008), {Tail reconnection triggering substorm onset.}, \textit{Science},
  \textit{321}(5891), 931--935, \doi{10.1126/science.1168045}.

\bibitem[{\textit{Asano et~al.}(2008)\textit{Asano, Nakamura, Shinohara,
  Fujimoto, Takada, Baumjohann, Owen, Fazakerley, Runov, Nagai, Lucek, and
  R{\`{e}}me}}]{Asano2008}
Asano, Y., R.~Nakamura, I.~Shinohara, M.~Fujimoto, T.~Takada, W.~Baumjohann,
  C.~J. Owen, a.~N. Fazakerley, A.~Runov, T.~Nagai, E.~a. Lucek, and
  H.~R{\`{e}}me (2008), {Electron flat-top distributions around the magnetic
  reconnection region}, \textit{J. Geophys. Res.}, \textit{113}(A1), A01,207,
  \doi{10.1029/2007JA012461}.

\bibitem[{\textit{Baker}(2002)}]{Baker2002}
Baker, D.~N. (2002), {Timing of magnetic reconnection initiation during a
  global magnetospheric substorm onset}, \doi{10.1029/2002GL015539}.

\bibitem[{\textit{Bessho et~al.}(2014)\textit{Bessho, Chen, Shuster, and
  Wang}}]{Bessho2015}
Bessho, N., L.~J. Chen, J.~R. Shuster, and S.~Wang (2014), {Electron
  distribution functions in the electron diffusion region of magnetic
  reconnection: Physics behind the fine structures}, \textit{Geophys. Res.
  Lett.}, \textit{41}(24), 8688--8695, \doi{10.1002/2014GL062034}.

\bibitem[{\textit{Borg et~al.}(2005)\textit{Borg, {\O}ieroset, Phan, Mozer,
  Pedersen, Mouikis, McFadden, Twitty, Balogh, and R{\`{e}}me}}]{Borg2005}
Borg, a.~L., M.~{\O}ieroset, T.~D. Phan, F.~S. Mozer, A.~Pedersen, C.~Mouikis,
  J.~P. McFadden, C.~Twitty, A.~Balogh, and H.~R{\`{e}}me (2005), {Cluster
  encounter of a magnetic reconnection diffusion region in the near-Earth
  magnetotail on September 19, 2003}, \textit{Geophys. Res. Lett.},
  \textit{32}(19), 32--35, \doi{10.1029/2005GL023794}.

\bibitem[{\textit{Chateau and Meyer-Vernet}(1989)}]{Chateau1989}
Chateau, Y.~F., and N.~Meyer-Vernet (1989), {Electrostatic noise in
  non-Maxwellian plasmas: “Flat-top” distribution function}, \textit{J.
  Geophys. Res.}, \textit{94}(A11), 15,407, \doi{10.1029/JA094iA11p15407}.

\bibitem[{\textit{Chen et~al.}(2008)\textit{Chen, Bessho, Lefebvre, Vaith,
  Fazakerley, Bhattacharjee, Puhl-Quinn, Runov, Khotyaintsev, Vaivads,
  Georgescu, and Torbert}}]{Chen2008}
Chen, L.~J., N.~Bessho, B.~Lefebvre, H.~Vaith, A.~Fazakerley, A.~Bhattacharjee,
  P.~a. Puhl-Quinn, A.~Runov, Y.~Khotyaintsev, A.~Vaivads, E.~Georgescu, and
  R.~Torbert (2008), {Evidence of an extended electron current sheet and its
  neighboring magnetic island during magnetotail reconnection}, \textit{J.
  Geophys. Res. Sp. Phys.}, \textit{113}(12), 1--19,
  \doi{10.1029/2008JA013385}.

\bibitem[{\textit{Chen et~al.}(2009)\textit{Chen, Bessho, Lefebvre, Vaith,
  Asnes, Santolik, Fazakerley, Puhl-Quinn, Bhattacharjee, Khotyaintsev, Daly,
  and Torbert}}]{Chen2009}
Chen, L.-J., N.~Bessho, B.~Lefebvre, H.~Vaith, A.~Asnes, O.~Santolik,
  A.~Fazakerley, P.~Puhl-Quinn, A.~Bhattacharjee, Y.~Khotyaintsev, P.~Daly, and
  R.~Torbert (2009), {Multispacecraft observations of the electron current
  sheet, neighboring magnetic islands, and electron acceleration during
  magnetotail reconnection}, \textit{Phys. Plasmas}, \textit{16}(5), 056,501,
  \doi{10.1063/1.3112744}.

\bibitem[{\textit{Daughton et~al.}(2006)\textit{Daughton, Scudder, and
  Karimabadi}}]{Daughton2006b}
Daughton, W., J.~D. Scudder, and H.~Karimabadi (2006), {Fully kinetic
  simulations of undriven magnetic reconnection with open boundary conditions},
  \textit{Phys. Plasmas}, \textit{13}(7), \doi{10.1063/1.2218817}.

\bibitem[{\textit{Drake et~al.}(2003)\textit{Drake, Swisdak, Cattell, Shay,
  Rogers, and Zeiler}}]{Drake2003}
Drake, J.~F., M.~Swisdak, C.~Cattell, M.~a. Shay, B.~N. Rogers, and a.~Zeiler
  (2003), {Formation of electron holes and particle energization during
  magnetic reconnection.}, \textit{Science}, \textit{299}(5608), 873--877,
  \doi{10.1126/science.1080333}.

\bibitem[{\textit{Drake et~al.}(2006)\textit{Drake, Swisdak, Che, and
  Shay}}]{Drake2006}
Drake, J.~F., M.~Swisdak, H.~Che, and M.~A. Shay (2006), {Electron acceleration
  from contracting magnetic islands during reconnection.}, \textit{Nature},
  \textit{443}(7111), 553--6, \doi{10.1038/nature05116}.

\bibitem[{\textit{Dum}(1978)}]{Dum1978}
Dum, C.~T. (1978), {Anomalous heating by ion sound turbulence}, \textit{Phys.
  Fluids}, \textit{21}(6), 945, \doi{10.1063/1.862338}.

\bibitem[{\textit{Eastwood et~al.}(2009)\textit{Eastwood, Phan, Bale, and
  Tjulin}}]{Eastwood2009}
Eastwood, J.~P., T.~D. Phan, S.~D. Bale, and A.~Tjulin (2009), {Observations of
  Turbulence Generated by Magnetic Reconnection}, \textit{Phys. Rev. Lett.},
  \textit{102}(3), 1--4, \doi{10.1103/PhysRevLett.102.035001}.

\bibitem[{\textit{Eastwood et~al.}(2010)\textit{Eastwood, Phan, {\O}ieroset,
  and Shay}}]{Eastwood2010a}
Eastwood, J.~P., T.~D. Phan, M.~{\O}ieroset, and M.~a. Shay (2010), {Average
  properties of the magnetic reconnection ion diffusion region in the Earth's
  magnetotail: The 2001–2005 Cluster observations and comparison with
  simulations}, \textit{J. Geophys. Res.}, \textit{115}(A8), A08,215,
  \doi{10.1029/2009JA014962}.

\bibitem[{\textit{Egedal et~al.}(2005)\textit{Egedal, {\O}ieroset, Fox, and
  Lin}}]{Egedal2005}
Egedal, J., M.~{\O}ieroset, W.~Fox, and R.~P. Lin (2005), {In Situ Discovery of
  an Electrostatic Potential, Trapping Electrons and Mediating Fast
  Reconnection in the Earth's Magnetotail}, \textit{Phys. Rev. Lett.},
  \textit{94}(2), 025,006, \doi{10.1103/PhysRevLett.94.025006}.

\bibitem[{\textit{Egedal et~al.}(2008)\textit{Egedal, Fox, Katz, Porkolab,
  {\O}ieroset, Lin, Daughton, and Drake}}]{Egedal2008}
Egedal, J., W.~Fox, N.~Katz, M.~Porkolab, M.~{\O}ieroset, R.~P. Lin,
  W.~Daughton, and J.~F. Drake (2008), {Evidence and theory for trapped
  electrons in guide field magnetotail reconnection}, \textit{J. Geophys.
  Res.}, \textit{113}(A12), A12,207, \doi{10.1029/2008JA013520}.

\bibitem[{\textit{Egedal et~al.}(2010)\textit{Egedal, L{\^{e}}, Katz, Chen,
  Lefebvre, Daughton, and FAZAKERLEY}}]{Egedal2010a}
Egedal, J., A.~L{\^{e}}, N.~Katz, L.~J. Chen, B.~Lefebvre, W.~Daughton, and
  A.~FAZAKERLEY (2010), {Cluster observations of bidirectional beams caused by
  electron trapping during antiparallel reconnection}, \textit{J. Geophys.
  Res.}, \textit{115}(A3), A03,214, \doi{10.1029/2009JA014650}.

\bibitem[{\textit{Egedal et~al.}(2013)\textit{Egedal, Le, and
  Daughton}}]{Egedal2013}
Egedal, J., A.~Le, and W.~Daughton (2013), {A review of pressure anisotropy
  caused by electron trapping in collisionless plasma, and its implications for
  magnetic reconnection}, \doi{10.1063/1.4811092}.

\bibitem[{\textit{Egedal et~al.}(2015)\textit{Egedal, Daughton, Le, and
  Borg}}]{Egedal2015}
Egedal, J., W.~Daughton, A.~Le, and a.~L. Borg (2015), {Double layer electric
  fields aiding the production of energetic flat-top distributions and
  superthermal electrons within magnetic reconnection exhausts}, \textit{Phys.
  Plasmas}, \textit{22}(10), 101,208, \doi{10.1063/1.4933055}.

\bibitem[{\textit{Feldman et~al.}(1982)\textit{Feldman, Bame, Gary, Gosling,
  McComas, Thomsen, Paschmann, Sckopke, Hoppe, and Russell}}]{Feldman1982}
Feldman, W.~C., S.~Bame, S.~Gary, J.~Gosling, D.~J. McComas, M.~F. Thomsen,
  G.~Paschmann, N.~Sckopke, M.~Hoppe, and C.~T. Russell (1982), {Electron
  Heating Within the Earth's Bow Shock}, \textit{Phys. Rev. Lett.},
  \textit{49}(3), 199--201, \doi{10.1103/PhysRevLett.49.199}.

\bibitem[{\textit{Feldman et~al.}(1983)\textit{Feldman, Anderson, Bame,
  Gosling, Zwickl, and Smith}}]{Feldman1983}
Feldman, W.~C., R.~C. Anderson, S.~J. Bame, J.~T. Gosling, R.~D. Zwickl, and
  E.~J. Smith (1983), {Electron velocity distributions near interplantary
  shocks}, \textit{J. Geophys. Res.}, \textit{88}(A12), 9949,
  \doi{10.1029/JA088iA12p09949}.

\bibitem[{\textit{Fu et~al.}(2006)\textit{Fu, Lu, and Wang}}]{Fu2006}
Fu, X.~R., Q.~M. Lu, and S.~Wang (2006), {The process of electron acceleration
  during collisionless magnetic reconnection}, \textit{Phys. Plasmas},
  \textit{13}(1), 012,309, \doi{10.1063/1.2164808}.

\bibitem[{\textit{Fujimoto}(2006)}]{Fujimoto2006c}
Fujimoto, K. (2006), {Time evolution of the electron diffusion region and the
  reconnection rate in fully kinetic and large system}, \textit{Phys. Plasmas},
  \textit{13}(7), \doi{10.1063/1.2220534}.

\bibitem[{\textit{Fujimoto}(2014)}]{Fujimoto2014}
Fujimoto, K. (2014), {Wave activities in separatrix regions of magnetic
  reconnection}, \textit{Geophys. Res. Lett.}, \textit{41}(8), 2721--2728,
  \doi{10.1002/2014GL059893}.

\bibitem[{\textit{Fujimoto et~al.}(1997)\textit{Fujimoto, Nakamura, Shinohara,
  Nagai, Mukai, Saito, Yamamoto, and Kokubun}}]{Fujimoto1997a}
Fujimoto, M., M.~S. Nakamura, I.~Shinohara, T.~Nagai, T.~Mukai, Y.~Saito,
  T.~Yamamoto, and S.~Kokubun (1997), {Observations of earthward streaming
  electrons at the trailing boundary of a plasmoid}, \textit{Geophys. Res.
  Lett.}, \textit{24}(22), 2893--2896, \doi{10.1029/97GL02821}.

\bibitem[{\textit{Hesse et~al.}(1999)\textit{Hesse, Schindler, Birn, and
  Kuznetsova}}]{Hesse1999}
Hesse, M., K.~Schindler, J.~Birn, and M.~Kuznetsova (1999), {The diffusion
  region in collisionless magnetic reconnection}, \textit{Phys. Plasmas},
  \textit{6}(5), 1781, \doi{10.1063/1.873436}.

\bibitem[{\textit{Hesse et~al.}(2014)\textit{Hesse, Aunai, Birn, Cassak,
  Denton, Drake, Gombosi, Hoshino, Matthaeus, Sibeck, and
  Zenitani}}]{Hesse2014}
Hesse, M., N.~Aunai, J.~Birn, P.~Cassak, R.~E. Denton, J.~F. Drake, T.~Gombosi,
  M.~Hoshino, W.~Matthaeus, D.~Sibeck, and S.~Zenitani (2014), {Theory and
  Modeling for the Magnetospheric Multiscale Mission}, \textit{Space Sci.
  Rev.}, \doi{10.1007/s11214-014-0078-y}.

\bibitem[{\textit{Hoshino}(2005)}]{Hoshino2005}
Hoshino, M. (2005), {Electron surfing acceleration in magnetic reconnection},
  \textit{J. Geophys. Res.}, \textit{110}(A10), A10,215,
  \doi{10.1029/2005JA011229}.

\bibitem[{\textit{Karimabadi et~al.}(2007)\textit{Karimabadi, Daughton, and
  Scudder}}]{Karimabadi2007}
Karimabadi, H., W.~Daughton, and J.~D. Scudder (2007), {Multi-scale structure
  of the electron diffusion region}, \textit{Geophys. Res. Lett.},
  \textit{34}(13), 1--5, \doi{10.1029/2007GL030306}.

\bibitem[{\textit{Livadiotis}(2016)}]{Livadiotis2016}
Livadiotis, G. (2016), {Non-Euclidean-normed Statistical Mechanics},
  \textit{Phys. A Stat. Mech. its Appl.}, \textit{445}, 240--255,
  \doi{10.1016/j.physa.2015.11.002}.

\bibitem[{\textit{Lu et~al.}(2010)\textit{Lu, Huang, Xie, Wang, Wu, Vaivads,
  and Wang}}]{Lu2010}
Lu, Q., C.~Huang, J.~Xie, R.~Wang, M.~Wu, A.~Vaivads, and S.~Wang (2010),
  {Features of separatrix regions in magnetic reconnection: Comparison of 2-D
  particle-in-cell simulations and Cluster observations}, \textit{J. Geophys.
  Res.}, \textit{115}(A11), A11,208, \doi{10.1029/2010JA015713}.

\bibitem[{\textit{Manapat et~al.}(2006)\textit{Manapat, {\O}ieroset, Phan, Lin,
  and Fujimoto}}]{Manapat2006}
Manapat, M., M.~{\O}ieroset, T.~D. Phan, R.~P. Lin, and M.~Fujimoto (2006),
  {Field-aligned electrons at the lobe/plasma sheet boundary in the
  mid-to-distant magnetotail and their association with reconnection},
  \textit{Geophys. Res. Lett.}, \textit{33}(5), L05,101,
  \doi{10.1029/2005GL024971}.

\bibitem[{\textit{Nagai et~al.}(2001)\textit{Nagai, Shinohara, Fujimoto,
  Hoshino, Saito, Machida, and Mukai}}]{Nagai2001a}
Nagai, T., I.~Shinohara, M.~Fujimoto, M.~Hoshino, Y.~Saito, S.~Machida, and
  T.~Mukai (2001), {Geotail observations of the Hall current system: Evidence
  of magnetic reconnection in the magnetotail}, \doi{10.1029/2001JA900038}.

\bibitem[{\textit{Nagai et~al.}(2011)\textit{Nagai, Shinohara, Fujimoto,
  Matsuoka, Saito, and Mukai}}]{Nagai2011}
Nagai, T., I.~Shinohara, M.~Fujimoto, A.~Matsuoka, Y.~Saito, and T.~Mukai
  (2011), {Construction of magnetic reconnection in the near-Earth magnetotail
  with Geotail}, \textit{J. Geophys. Res.}, \textit{116}, A04,222,
  \doi{10.1029/2010JA016283}.

\bibitem[{\textit{Nagai et~al.}(2013)\textit{Nagai, Zenitani, Shinohara,
  Nakamura, Fujimoto, Saito, and Mukai}}]{Nagai2013}
Nagai, T., S.~Zenitani, I.~Shinohara, R.~Nakamura, M.~Fujimoto, Y.~Saito, and
  T.~Mukai (2013), {Ion and electron dynamics in the ion-electron decoupling
  region of magnetic reconnection with Geotail observations}, \textit{J.
  Geophys. Res. Sp. Phys.}, \textit{118}(12), 7703--7713,
  \doi{10.1002/2013JA019135}.

\bibitem[{\textit{Ng et~al.}(2011)\textit{Ng, Egedal, Le, Daughton, and
  Chen}}]{Ng2011}
Ng, J., J.~Egedal, A.~Le, W.~Daughton, and L.~J. Chen (2011), {Kinetic
  structure of the electron diffusion region in antiparallel magnetic
  reconnection}, \textit{Phys. Rev. Lett.}, \textit{106}(FEBRUARY), 1--4,
  \doi{10.1103/PhysRevLett.106.065002}.

\bibitem[{\textit{Ng et~al.}(2012)\textit{Ng, Egedal, Le, and
  Daughton}}]{Ng2012}
Ng, J., J.~Egedal, A.~Le, and W.~Daughton (2012), {Phase space structure of the
  electron diffusion region in reconnection with weak guide fields},
  \textit{Phys. Plasmas}, \textit{19}(2012), \doi{10.1063/1.4766895}.

\bibitem[{\textit{{\O}ieroset et~al.}(2001)\textit{{\O}ieroset, Phan, Fujimoto,
  Lin, and Lepping}}]{Oieroset2001}
{\O}ieroset, M., T.~D. Phan, M.~Fujimoto, R.~P. Lin, and R.~P. Lepping (2001),
  {In situ detection of collisionless reconnection in the Earth's magnetotail},
  \textit{Nature}, \textit{412}(July), 414--417.

\bibitem[{\textit{Oka et~al.}(2008)\textit{Oka, Fujimoto, Nakamura, Shinohara,
  and Nishikawa}}]{Oka2008}
Oka, M., M.~Fujimoto, T.~Nakamura, I.~Shinohara, and K.-I. Nishikawa (2008),
  {Magnetic Reconnection by a Self-Retreating X Line}, \textit{Phys. Rev.
  Lett.}, \textit{101}(20), 205,004, \doi{10.1103/PhysRevLett.101.205004}.

\bibitem[{\textit{Oka et~al.}(2010{\natexlab{a}})\textit{Oka, Phan, Krucker,
  Fujimoto, and Shinohara}}]{Oka2010}
Oka, M., T.~D. Phan, S.~Krucker, M.~Fujimoto, and I.~Shinohara
  (2010{\natexlab{a}}), {ELECTRON ACCELERATION BY MULTI-ISLAND COALESCENCE},
  \textit{Astrophys. J.}, \textit{714}(1), 915--926,
  \doi{10.1088/0004-637X/714/1/915}.

\bibitem[{\textit{Oka et~al.}(2010{\natexlab{b}})\textit{Oka, Fujimoto,
  Shinohara, and Phan}}]{Oka2010b}
Oka, M., M.~Fujimoto, I.~Shinohara, and T.~D. Phan (2010{\natexlab{b}}),
  {“Island surfing” mechanism of electron acceleration during magnetic
  reconnection}, \textit{J. Geophys. Res.}, \textit{115}(A8), A08,223,
  \doi{10.1029/2010JA015392}.

\bibitem[{\textit{Oka et~al.}(2011)\textit{Oka, Phan, Eastwood, Angelopoulos,
  Murphy, {\O}ieroset, Miyashita, Fujimoto, Mcfadden, and Larson}}]{oka2011}
Oka, M., T.~D. Phan, J.~P. Eastwood, V.~Angelopoulos, N.~A. Murphy,
  M.~{\O}ieroset, Y.~Miyashita, M.~Fujimoto, J.~P. Mcfadden, and D.~Larson
  (2011), {Magnetic Reconnection X-Line Retreat Associated with Dipolarization
  of the Earth's Magnetosphere}, \textit{Geophys. Res. Lett.}, \textit{38},
  L20,105.

\bibitem[{\textit{Olbert}(1968)}]{Olbert1968}
Olbert, S. (1968), {Summary of experimental results from M.I.T. detector on
  IMP-1.}, in \textit{Phys. Magnetos.}, edited by R.~Carovillano, J.~F. McClay,
  and H.~R. Radoski, pp. 641--659, Springer Netherlands, Dordrecht, Holland,
  \doi{10.1007/978-94-010-3467-8{\_}23}.

\bibitem[{\textit{Pedersen et~al.}(2001)\textit{Pedersen, D{\'{e}}cr{\'{e}}au,
  Escoubet, Gustafsson, Laakso, Lindqvist, Lybekk, Masson, Mozer, and
  Vaivads}}]{Pedersen2001}
Pedersen, A., P.~D{\'{e}}cr{\'{e}}au, C.-P. Escoubet, G.~Gustafsson, H.~Laakso,
  P.-a. Lindqvist, B.~Lybekk, A.~Masson, F.~S. Mozer, and A.~Vaivads (2001),
  {Four-point high time resolution information on electron densities by the
  electric field experiments (EFW) on Cluster}, \textit{Ann. Geophys.},
  \textit{19}(10/12), 1483--1489, \doi{10.5194/angeo-19-1483-2001}.

\bibitem[{\textit{Pedersen et~al.}(2008)\textit{Pedersen, Lybekk, Andr{\'{e}},
  Eriksson, Masson, Mozer, Lindqvist, D{\'{e}}cr{\'{e}}au, Dandouras, Sauvaud,
  Fazakerley, Taylor, Paschmann, Svenes, Torkar, and Whipple}}]{Pedersen2008}
Pedersen, A., B.~Lybekk, M.~Andr{\'{e}}, A.~Eriksson, A.~Masson, F.~S. Mozer,
  P.-A. Lindqvist, P.~M.~E. D{\'{e}}cr{\'{e}}au, I.~Dandouras, J.-A. Sauvaud,
  A.~Fazakerley, M.~Taylor, G.~Paschmann, K.~R. Svenes, K.~Torkar, and
  E.~Whipple (2008), {Electron density estimations derived from spacecraft
  potential measurements on Cluster in tenuous plasma regions}, \textit{J.
  Geophys. Res.}, \textit{113}(A7), A07S33, \doi{10.1029/2007JA012636}.

\bibitem[{\textit{Petschek}(1984)}]{Petschek1984}
Petschek, H.~E. (1984), {Magnetic Field Annihilation}, \textit{Phys. Sol.
  Flares, Proc. AAS-NASA Symp.}, p. 425.

\bibitem[{\textit{Phan et~al.}(2007)\textit{Phan, Drake, Shay, Mozer, and
  Eastwood}}]{Phan2007a}
Phan, T.~D., J.~F. Drake, M.~A. Shay, F.~S. Mozer, and J.~P. Eastwood (2007),
  {Evidence for an Elongated (>60 Ion Skin Depths) Electron Diffusion Region
  during Fast Magnetic Reconnection}, \textit{Phys. Rev. Lett.}, \textit{99},
  255,002, \doi{10.1103/PhysRevLett.99.255002}.

\bibitem[{\textit{Pritchett}(2006)}]{Pritchett2006a}
Pritchett, P.~L. (2006), {Relativistic electron production during driven
  magnetic reconnection}, \textit{Geophys. Res. Lett.}, \textit{33}(13), 1--4,
  \doi{10.1029/2005GL025267}.

\bibitem[{\textit{Runov}(2003)}]{Runov2003c}
Runov, a. (2003), {Current sheet structure near magnetic X-line observed by
  Cluster}, \doi{10.1029/2002GL016730}.

\bibitem[{\textit{Scudder et~al.}(2012)\textit{Scudder, Holdaway, Daughton,
  Karimabadi, Roytershteyn, Russell, and Lopez}}]{Scudder2012}
Scudder, J.~D., R.~D. Holdaway, W.~Daughton, H.~Karimabadi, V.~Roytershteyn,
  C.~T. Russell, and J.~Y. Lopez (2012), {First resolved observations of the
  demagnetized electron-diffusion region of an astrophysical
  magnetic-reconnection site}, \textit{Phys. Rev. Lett.}, \textit{108}(22),
  1--5, \doi{10.1103/PhysRevLett.108.225005}.

\bibitem[{\textit{Shay et~al.}(2001)\textit{Shay, Drake, Rogers, and
  Denton}}]{Shay2001}
Shay, M.~a., J.~F. Drake, B.~N. Rogers, and R.~E. Denton (2001),
  {Alfv{\'{e}}nic collisionless magnetic reconnection and the Hall term},
  \textit{J. Geophys. Res.}, \textit{106}(A3), 3759--3772,
  \doi{10.1029/1999JA001007}.

\bibitem[{\textit{Shay et~al.}(2007)\textit{Shay, Drake, and
  Swisdak}}]{Shay2007}
Shay, M.~A., J.~F. Drake, and M.~Swisdak (2007), {Two-Scale Structure of the
  Electron Dissipation Region during Collisionless Magnetic Reconnection},
  \textit{Phys. Rev. Lett.}, \textit{99}, 155,002,
  \doi{10.1103/PhysRevLett.99.155002}.

\bibitem[{\textit{Shuster et~al.}(2015)\textit{Shuster, Chen, Hesse, Argall,
  Daughton, Torbert, and Bessho}}]{Shuster2015}
Shuster, J.~R., L.~J. Chen, M.~Hesse, M.~R. Argall, W.~Daughton, R.~B. Torbert,
  and N.~Bessho (2015), {Spatiotemporal evolution of electron characteristics
  in the electron diffusion region of magnetic reconnection: Implications for
  acceleration and heating}, \textit{Geophys. Res. Lett.}, \textit{42}(8),
  2586--2593, \doi{10.1002/2015GL063601}.

\bibitem[{\textit{Sibeck et~al.}(1985)\textit{Sibeck, Siscoe, Slavin, Smith,
  Tsurutani, and Lepping}}]{Sibeck1985}
Sibeck, D.~G., G.~L. Siscoe, J.~A. Slavin, E.~J. Smith, B.~T. Tsurutani, and
  R.~P. Lepping (1985), {The distant magnetotail's response to a strong
  interplanetary magnetic field B y : Twisting, flattening, and field line
  bending}, \textit{J. Geophys. Res.}, \textit{90}(A5), 4011,
  \doi{10.1029/JA090iA05p04011}.

\bibitem[{\textit{Sonnerup}(1979)}]{Sonnerup1979}
Sonnerup, B. U.~O. (1979), {Magnetic field reconnection}, in \textit{Sol. Syst.
  Plasma Phys. Vol. III}, edited by L.~T. Lanzerotti, C.~F. Kennel, and E.~N.
  Parker, pp. 47--108, North-Holland, Amsterdam.

\bibitem[{\textit{Tanaka et~al.}(2010)\textit{Tanaka, Yumura, Fujimoto,
  Shinohara, Badman, and Grocott}}]{Tanaka2010}
Tanaka, K.~G., T.~Yumura, M.~Fujimoto, I.~Shinohara, S.~V. Badman, and
  A.~Grocott (2010), {Merging of magnetic islands as an efficient accelerator
  of electrons}, \textit{Phys. Plasmas}, \textit{17}(2010),
  \doi{10.1063/1.3491123}.

\bibitem[{\textit{Teh et~al.}(2012)\textit{Teh, Nakamura, Fujimoto, Kronberg,
  Fazakerley, Daly, and Baumjohann}}]{Teh2012a}
Teh, W.-L., R.~Nakamura, M.~Fujimoto, E.~a. Kronberg, a.~N. Fazakerley, P.~W.
  Daly, and W.~Baumjohann (2012), {Electron dynamics in the reconnection ion
  diffusion region}, \textit{J. Geophys. Res. Sp. Phys.}, \textit{117}(A12),
  n/a--n/a, \doi{10.1029/2012JA017896}.

\bibitem[{\textit{Terasawa}(1983)}]{Terasawa1983a}
Terasawa, T. (1983), {Hall current effect on tearing mode instability},
  \textit{Geophys. Res. Lett.}, \textit{10}(6), 475--478,
  \doi{10.1029/GL010i006p00475}.

\bibitem[{\textit{Thomsen et~al.}(1983)\textit{Thomsen, Barr, Gary, Feldman,
  and Cole}}]{Thomsen1983a}
Thomsen, M.~F., H.~C. Barr, S.~P. Gary, W.~C. Feldman, and T.~E. Cole (1983),
  {Stability of electron distributions within the Earth's bow shock},
  \textit{J. Geophys. Res.}, \textit{88}(A4), 3035,
  \doi{10.1029/JA088iA04p03035}.

\bibitem[{\textit{Tsuneta and Naito}(1998)}]{Tsuneta1998a}
Tsuneta, S., and T.~Naito (1998), {Fermi Acceleration at the Fast Shock in a
  Solar Flare and the Impulsive Loop-Top Hard X-Ray Source}, \textit{Astrophys.
  J.}, \textit{495}(1), L67--L70, \doi{10.1086/311207}.

\bibitem[{\textit{Tsyganenko}(1995)}]{Tsyganenko1995a}
Tsyganenko, N.~a. (1995), {Modeling the Earth's magnetospheric magnetic field
  confined within a realistic magnetopause}, \textit{J. Geophys. Res.},
  \textit{100}(A4), 5599--5612, \doi{10.1029/94JA03193}.

\bibitem[{\textit{Tsyganenko}(2004)}]{Tsyganenko2004}
Tsyganenko, N.~a. (2004), {Global shape of the magnetotail current sheet as
  derived from Geotail and Polar data}, \textit{J. Geophys. Res.},
  \textit{109}(A3), A03,218, \doi{10.1029/2003JA010062}.

\bibitem[{\textit{Vasyliunas}(1968)}]{Vasyliunas1968a}
Vasyliunas, V.~M. (1968), {A survey of low-energy electrons in the evening
  sector of the magnetosphere with OGO 1 and OGO 3}, \textit{J. Geophys. Res.},
  \textit{73}(9), 2839--2884, \doi{10.1029/JA073i009p02839}.

\bibitem[{\textit{Wang et~al.}(2010{\natexlab{a}})\textit{Wang, Lu, Du, and
  Wang}}]{Wang2010a}
Wang, R., Q.~Lu, A.~Du, and S.~Wang (2010{\natexlab{a}}), {In Situ Observations
  of a Secondary Magnetic Island in an Ion Diffusion Region and Associated
  Energetic Electrons}, \textit{Phys. Rev. Lett.}, \textit{104}(17), 175,003,
  \doi{10.1103/PhysRevLett.104.175003}.

\bibitem[{\textit{Wang et~al.}(2010{\natexlab{b}})\textit{Wang, Lu, Li, Huang,
  and Wang}}]{Wang2010b}
Wang, R., Q.~Lu, X.~Li, C.~Huang, and S.~Wang (2010{\natexlab{b}}),
  {Observations of energetic electrons up to 200 keV associated with a
  secondary island near the center of an ion diffusion region: A Cluster case
  study}, \textit{J. Geophys. Res.}, \textit{115}(A11), A11,201,
  \doi{10.1029/2010JA015473}.

\bibitem[{\textit{Wang et~al.}(2014)\textit{Wang, Lu, Khotyaintsev, Volwerk,
  Du, Nakamura, Gonzalez, Sun, Baumjohann, Li, Zhang, Fazakerley, Huang, and
  Wu}}]{Wang2014}
Wang, R., Q.~Lu, Y.~V. Khotyaintsev, M.~Volwerk, A.~Du, R.~Nakamura, W.~D.
  Gonzalez, X.~Sun, W.~Baumjohann, X.~Li, T.~Zhang, A.~N. Fazakerley, C.~Huang,
  and M.~Wu (2014), {Observation of double layer in the separatrix region
  during magnetic reconnection}, \textit{Geophys. Res. Lett.}, \textit{41}(14),
  4851--4858, \doi{10.1002/2014GL061157}.

\bibitem[{\textit{Zenitani et~al.}(2012)\textit{Zenitani, Shinohara, and
  Nagai}}]{Zenitani2012a}
Zenitani, S., I.~Shinohara, and T.~Nagai (2012), {Evidence for the dissipation
  region in magnetotail reconnection}, \textit{Geophys. Res. Lett.},
  \textit{39}(11), n/a--n/a, \doi{10.1029/2012GL051938}.

\end{thebibliography}


%
%
\end{article}
%
%
%
%
%
%
%
%


\end{document}